\documentclass{PoS}

\def\bra#1#2{\ifx#2\ket\langle#1\else\langle#1\vert\fi#2}
\def\ket#1{\vert#1\rangle}

\leftline{\parbox{3cm}{\large\rm HU-EP-08/66 LU-ITP 2008/04}
\vspace{1cm}}

\title{Higher-loop gluon and ghost propagators \\ in Landau gauge  
from numerical stochastic perturbation theory }

\ShortTitle{Higher-loop gluon and ghost propagators in Landau gauge from
NSPT}

\author{F.~Di~Renzo\\
Universit\`a di Parma \& INFN, Viale Usberti 7/A, I-43100 Parma, Italy\\
E-mail: \email{francesco.direnzo@fis.unipr.it}}

\author{E.-M.~Ilgenfritz\\
Institut f\"ur Physik, Humboldt-Universit\"at zu Berlin, Newtonstr. 15, D-12489 Berlin, Germany \\
Institut f\"ur Physik, Karl-Franzens-Universit\"at Graz, Universit\"atsplatz 5, A-8010 Graz, Austria \\
E-mail: \email{ilgenfri@physik.hu-berlin.de}}

\author{H.~Perlt\\
Institut f\"ur Theoretische Physik, Universit\"at Leipzig, PF 100 920, D-04009 Leipzig, Germany\\
E-mail: \email{Holger.Perlt@itp.uni-leipzig.de}}

\author{\speaker{A.~Schiller}\\
Institut f\"ur Theoretische Physik, Universit\"at Leipzig, PF 100 920, D-04009 Leipzig, Germany\\
E-mail: \email{Arwed.Schiller@itp.uni-leipzig.de}}

\author{C.~Torrero\\
Institut f\"ur Theoretische Physik, Universit\"at Regensburg, Universit\"atsstr. 31, \\ D-93053 Regensburg, Germany\\
E-mail: \email{christian.torrero@physik.uni-regensburg.de}}

\abstract{
We present higher loop results for the gluon and ghost propagator in Landau
gauge on the lattice calculated in numerical stochastic perturbation theory.
We make predictions for the perturbative content of those propagators 
as function of the lattice momenta for finite lattices.
To find out their nonperturbative contributions, the logarithmic definition of the gauge fields 
and the corresponding Faddeev-Popov operator have to be implemented in the Monte Carlo simulations.
}
\FullConference{8th Conference Quark Confinement and the Hadron Spectrum\\
		 September 1-6, 2008\\
		 Mainz. Germany}

\begin{document}

\section{The Langevin equation and NSPT}
\label{sec:basics}

A typical task for Lattice Perturbation Theory (LPT) is the determination of renormalization 
factors or the separation of non-perturbative effects from observables related to confinement. 
The gluon and ghost propagators belong to this class.\footnote{See the talk 
by A. Maas at this conference~\cite{Maas:2008uz} trying to identify the 
non-perturbative content.}
For that higher-loop calculations are needed. Unfortunately, the diagrammatic approach is much 
more involved than in the continuum case. 
It is unlikely to go beyond the two-loop level in near future.

A promising alternative to diagrammtic LPT has been proven to be Numerical Stochastic Perturbation
Theory (NSPT). For a review see Ref.~\cite{DiRenzo:2004ge}. It makes it possible to obtain 
higher-loop results without computing vast numbers of Feynman diagrams.
Various applications of NSPT have been described in the past, see 
{\it e.g.}~\cite{plaq,Di Renzo:2004xn,DiRenzo:2006wd,DiRenzo:2006nh}.
Here we report on 
first NSPT studies of the Yang-Mills gluon and ghost propagator in Landau gauge. For other preparatory
results see also~\cite{Ilgenfritz:2007qj,DiRenzo:2007qf}.

NSPT is based on stochastic quantization, {\it i.e.}, a lattice variant of the Langevin equation. 
This equation describes the evolution of 4D fields with respect to a fictitious 
time $t$ under the influence of Gaussian random noise. In the limit $t \to \infty$ 
the whole set of gauge links $U$ is distributed according to Gibbs measure $\rm{exp}(-S_G[U])$.
In our simulations we use the Wilson gauge action $S_G[U]$, but an improved action would not
be much more complicated.

Other than in usual Langevin simulation, in NSPT the gauge link fields are expanded in powers 
of the bare coupling $g \propto \beta^{-1/2}$. Discretizing the Langevin time as $t = n \tau$, 
rescaling $\tau= \varepsilon / \beta$ and using the Euler scheme, a set of coupled equations 
emerges corresponding to different orders in the coupling constant.

From the resulting fields the  Green functions of interest can be constructed order by order in LPT.
In NSPT the algebra-valued gauge potential $A_{x,\mu}$ 
are related to the gauge lattice link fields $U_{x,\mu}$ by
\begin{equation}
  A_{x,\mu}= \log U_{x,\mu} \, . 
  \label{eq:Adef}
\end{equation}
Its expansion is given in the form
\begin{equation}
  A_{x,\mu} \to \sum_{l>0} \beta^{-l/2} A_{x,\mu}^{(l)} \, , \qquad
  A_{x,\mu}^{(l)}= T^a A_{x,\mu}^{a,(l)} \, .
  \label{eq:A-expansion}
\end{equation}
Each simultaneous Langevin update of the expansion coefficients $A_{x,\mu}^{(l)}$ is augmented
by a stochastic gauge-fixing step and by subtracting zero modes from $A^{(l)}$
as described in Ref.~\cite{DiRenzo:2004ge}.

In this talk we present perturbative contributions to the lattice gluon and ghost propagators 
in {\it minimal} Landau gauge. Before taking measurements, this gauge is achieved by iterative 
gauge transformations. We use a perturbatively expanded version of the Fourier-accelerated 
gauge-fixing method~\cite{Davies:1987vs}.

\section{The gluon and ghost propagators in NSPT}
\label{sec:NSPT-ghost}

The lattice gluon propagator $D^{ab}_{\mu\nu}(\hat{q})$ is the Fourier transform
of the gluon two-point function, {\it i.e.}, the expectation value
\begin{equation}
  D^{ab}_{\mu\nu}(\hat{q}) = \left\langle \widetilde{A}^a_{\mu}(k)
  \widetilde{A}^b_{\nu}(-k) \right\rangle = \delta^{ab} D_{\mu\nu}(\hat{q}) \, .
\label{eq:D-definition}
\end{equation}
$\widetilde{A}^a_{\mu}(k)$ is the Fourier transform of $A^a_{x,\mu}$, and $\hat{q}$ denotes
the physical discrete momentum corresponding to the integers $k_{\mu} \in \left(-L_{\mu}/2, L_{\mu}/2\right]$, 
\begin{equation}
  \hat{q}_{\mu}(k_{\mu}) = \frac{2}{a} \sin\left(\frac{\pi
      k_{\mu}}{L_{\mu}}\right)= \frac{2}{a} \sin\left(\frac{ a q_\mu}{2}\right) \, .
\label{eq:p-definition}
\end{equation}
In NSPT the different loop orders $n$  (even orders in $\beta^{-1/2}$) are constructed directly from
gauge fields $\widetilde{A}^{a,(l)}_\mu(k)$
\begin{equation}
  \delta^{ab} D_{\mu\nu}^{(n)}(\hat q) = \left\langle \,
  \sum_{l=1}^{2n+1}
  \left[ \widetilde{A}^{a,(l)}_{\mu}(k) \,
  \widetilde{A}^{b,(2n+2-l)}_{\nu}(-k) \right] \,
  \right\rangle \,.
\label{eq:Dn}
\end{equation}
Already the tree level term $D_{\mu\nu}^{(0)}$ results from
quantum fluctuations of gauge fields with $l=1$. 
Terms with non-integer $n=1/2,3/2,\dots$ in (\ref{eq:Dn}) have to vanish 
numerically.
Motivated by the structure of the continuum propagator in Landau gauge
we consider the so called gluon dressing function of different loop orders $n$
\begin{equation}
  \hat{J}_{\rm gl}^{(n)}(\hat q)=  \hat q^2 D^{(n)} (\hat q) \equiv  
  \frac{\hat q^2}{3} \sum_{\mu=1}^4 D_{\mu\mu}^{(n)}(\hat{q}) \, . 
\end{equation}

The ghost propagator is nothing but the inverse of the Faddeev-Popov (FP) operator 
$M$ which can be constructed in Landau gauge using the lattice covariant and left 
partial derivatives. Since the progagator is color diagonal, $G^{ab}(\hat q) = \delta^{ab} G(\hat q)$,  
it is obtained as the color trace
\begin{equation}
  G(\hat q  )= 
  \frac{1}{N_c^2-1} \left\langle {\rm{Tr_{adj}}}~M^{-1}(k)\right\rangle_U \,.
  \label{eq:G-definition}
\end{equation}
In (\ref{eq:G-definition}) $M^{-1}(k)$ is the Fourier transform of the inverse 
FP operator in real space. 
The perturbative expansion is based on the mapping
$
  \{A_{x,\mu}^{(l)} \} \,\rightarrow\, \{M^{(l)} \} \,\rightarrow\, \{\left[M^{-1}\right]^{\!(l)}\}
$.
Since $M$ is expanded in terms of $M^{(l)}$ (containing $A^{(l)}$), 
a {\sl recursive} inversion is possible:
\begin{eqnarray}
  \left[M^{~\!\!-1}\right]^{\!(0)} = \left[M^{~\!\!(0)}\right]^{\!-1} \,, \quad
  \left[M^{~\!\!-1}\right]^{(l)} = -\left[M^{~\!0}\right]^{\!-1}\;\sum_{j=0}^{l-1} M^{~\!(l-j)}\;
  \left[M^{~\!\!-1}\right]^{\!(j)} \, .
  \label{eq:inversion}
\end{eqnarray}
The momentum-space ghost propagator at $n$-loop order is obtained from even orders $l = 2 n$ of $M^{-1}$, 
sandwiching (\ref{eq:inversion}) between the plane-wave vectors $\ket {k,c}$ 
\begin{equation}
   G^{~\!\!(n)}(\hat q (k) ) = \frac{1}{N_c^2-1} \sum_c \langle k,c |\left[M^{~\!-1}\right]^{(l=2n)}| k,c \rangle \, . 
\end{equation}
Again non-integer $n$ (odd $l$) orders have to vanish within numerical precision.
We present our results in terms of the ghost dressing function: 
\begin{equation}
  \hat J_{\rm gh}^{~\!\!(n)}(\hat q) = \hat q^2 \; G^{(n)}(\hat q) \, . 
\end{equation}
Note that the perturbative construction of $M$ in terms of the gauge fields $A$ (and therefore
the corresponding ghost propagator) differ from the definition adopted in most Monte Carlo calculations.
For each chosen momentum 
$(k_1,k_2,k_3,k_4)$ and different colors $c$ of the plane wave $\ket {k,c}$ the propagator has to be 
calculated individually. This makes the measurement expensive.

\section{Some numerical results}
\label{sec:results}

The  configuration sequence of all $A^{(l)}$ created at finite $\varepsilon$
can be used to measure the perturbatively expanded observables.
Already at finite $\varepsilon$ the non-integer $n$ contributions to the
dressing functions have to vanish. 
We have studied the step size limit $\varepsilon \to 0$ working at $\varepsilon=0.07 \ldots 0.01$.
In order to make contact with standard 
infinite-volume LPT at vanishing lattice spacing the limits $L \to \infty$ and $a q\to 0$ have 
to be performed additionally.

\begin{figure}[!htb]
  \begin{center}
    \begin{tabular}{cc}
       \includegraphics[scale=0.59,clip=true] {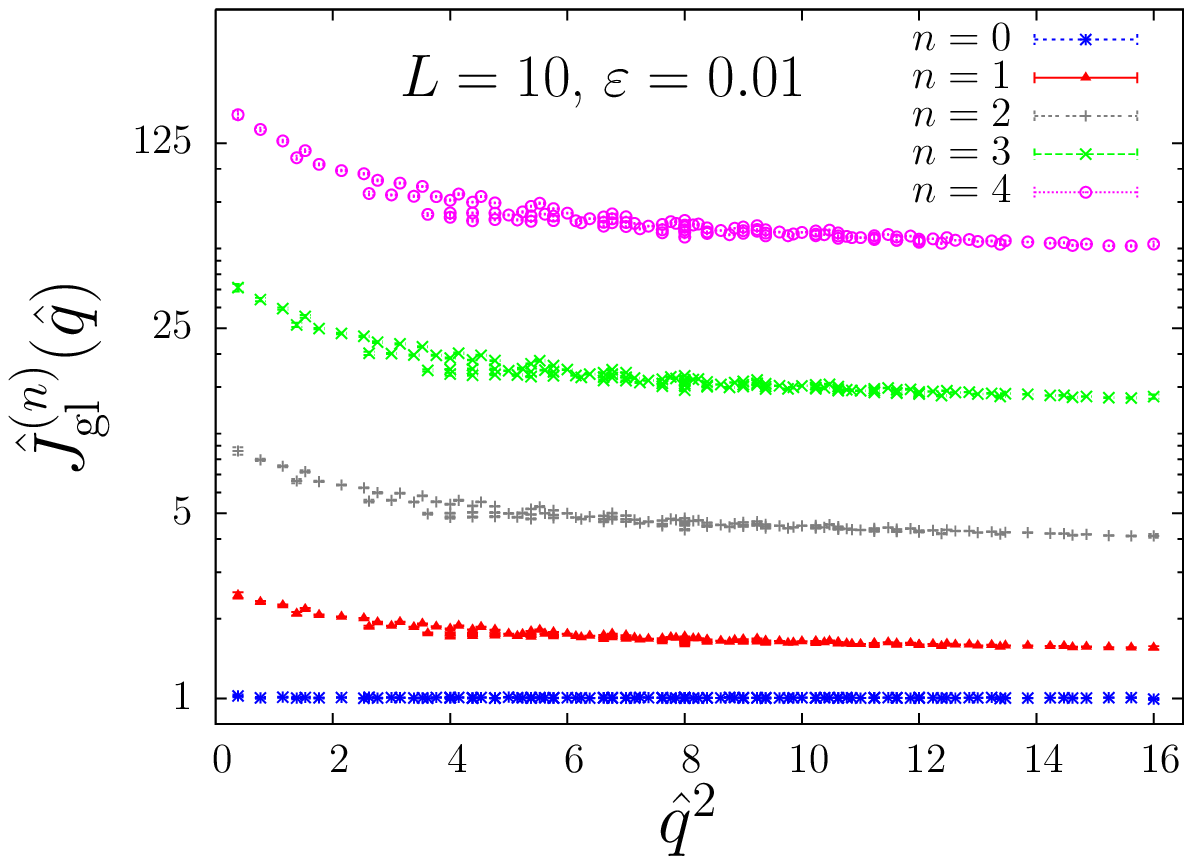}&
       \includegraphics[scale=0.59,clip=true] {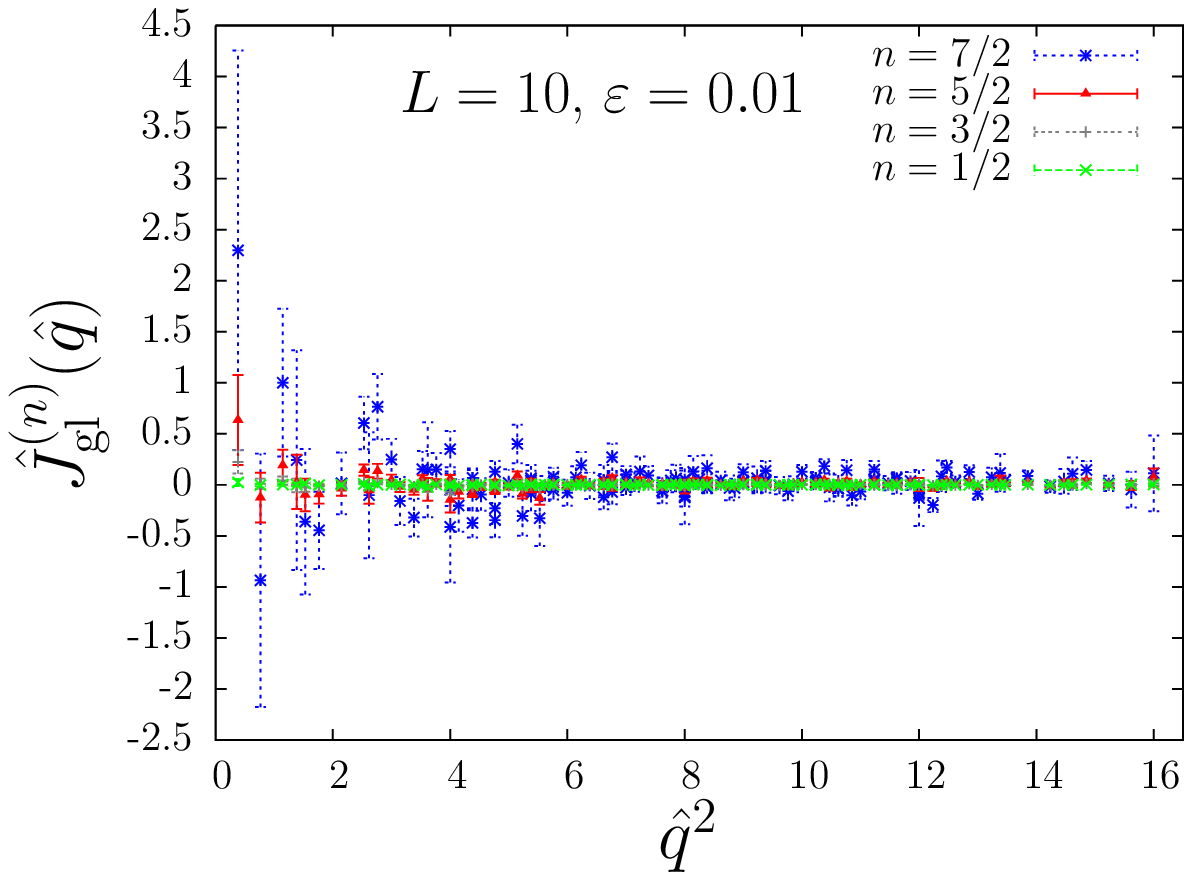} \\
    \end{tabular}
  \end{center}
  \vspace{-5mm}
  \caption{Measured gluon dressing function $\hat{J}_{\rm gl}(\hat q)$ vs. $\hat q^2$.
           Left: Separate loop contributions $\hat J_{\rm gl}^{(n)}(\hat q)$  vs. $\hat q^2$
           (for inequivalent lattice momentum 4-tuples) at $L=10$ and $\varepsilon=0.01$;
           right: the vanishing coefficient of the contribution $\propto \beta^{-n/2}$.}
  \label{fig:gluon}
\end{figure}

\begin{figure}[!htb]
  \begin{center}
    \begin{tabular}{cc}
       \includegraphics[scale=0.59,clip=true] {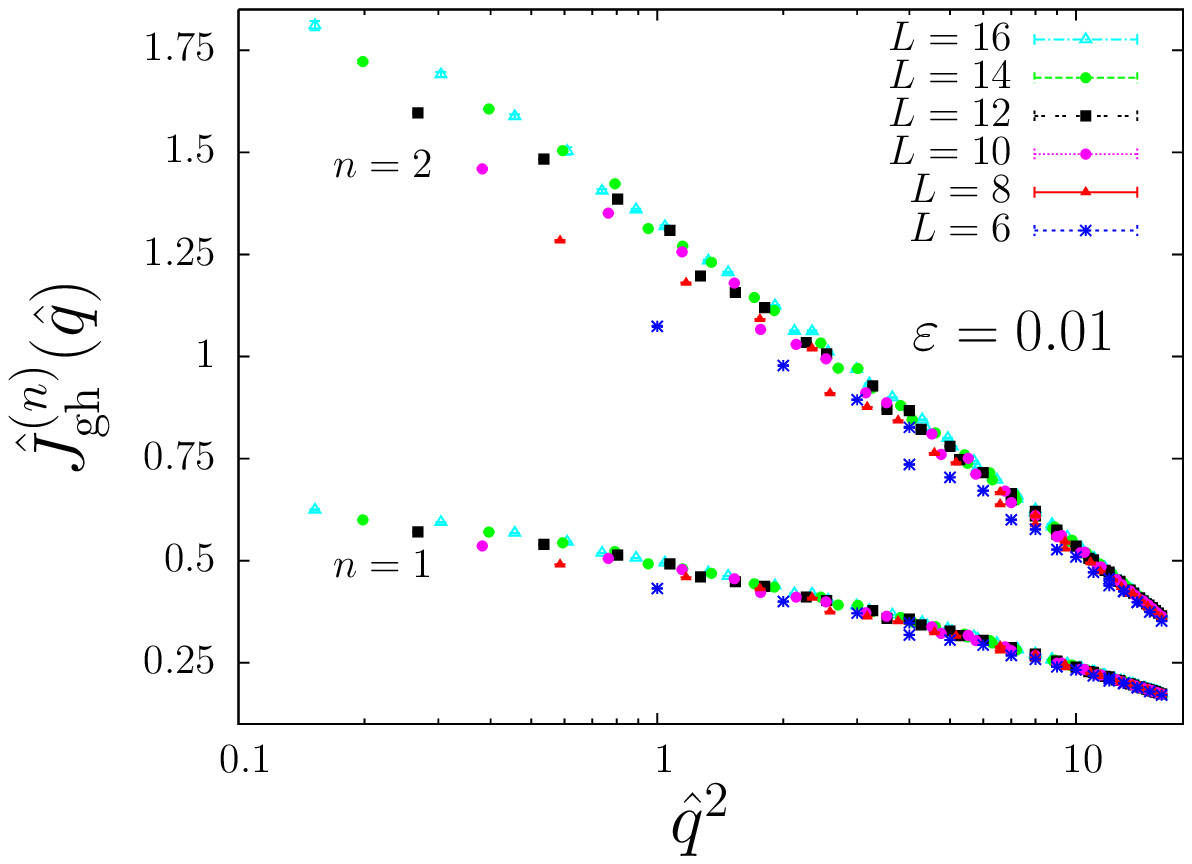}&
       \includegraphics[scale=0.59,clip=true] {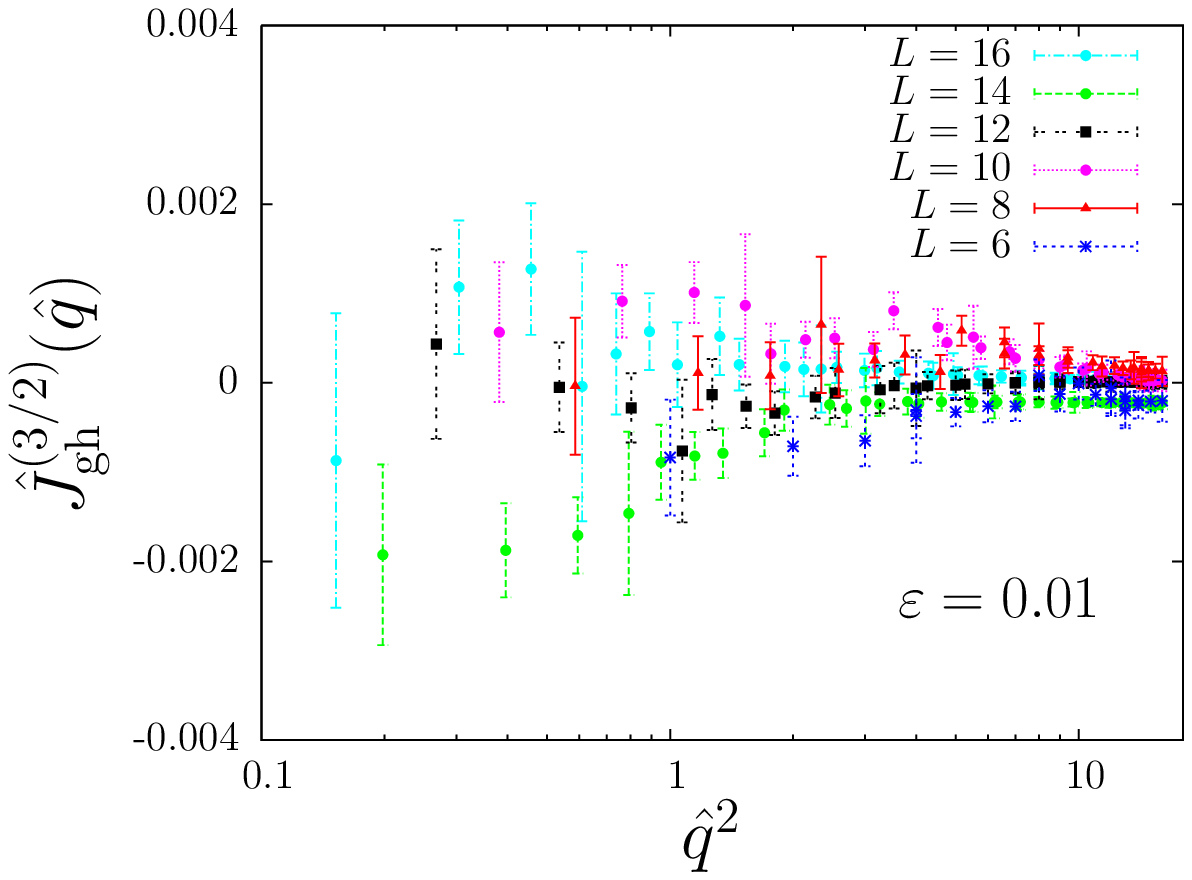} \\
    \end{tabular}
  \end{center}
  \vspace{-5mm}
  \caption{Measured ghost dressing function $\hat{J}_{\rm gh}(\hat q)$ vs. $\hat q^2$ 
           (for inequivalent lattice momentum 4-tuples close to
           the diagonal) for lattice sizes $L=6,\ldots,16$ and for the time step 
           $\varepsilon=0.01$.
           Left: The one-loop and two-loop contributions; right: the vanishing 
           coefficient of the contribution $\propto \beta^{-3/2}$.}
\label{fig:one_and_two_loop}
\end{figure}

In Fig.~\ref{fig:gluon} we present different orders of the gluon dressing function at the smallest 
time step $\varepsilon=0.01$ and lattice size $10^4$. The loop contributions are labelled by integer $n$, 
dressing contributions with noninteger $n$ vanish.
As an example of the ghost propagator we show in Fig.~\ref{fig:one_and_two_loop} 
the one- and two-loop results $\hat{J}_{\rm gh}^{~\!(1)}$ and $\hat{J}_{\rm gh}^{~\!(2)}$ 
for the dressing function together with the vanishing $\hat{J}_{\rm gh}^{~\!(l/2=3/2)}$.

For each set of inequivalent lattice momenta $(k_1,k_2,k_3,k_4)$ we have performed the
extrapolation to zero time step. Performing that limit, the individual loop contributions 
to the dressing functions are available for all lattice momenta at a given lattice size.
They can be compared with Monte Carlo results on finite lattices.

\begin{figure}[!htb]
  \begin{tabular}{cc}
    \includegraphics[scale=0.59,clip=true]{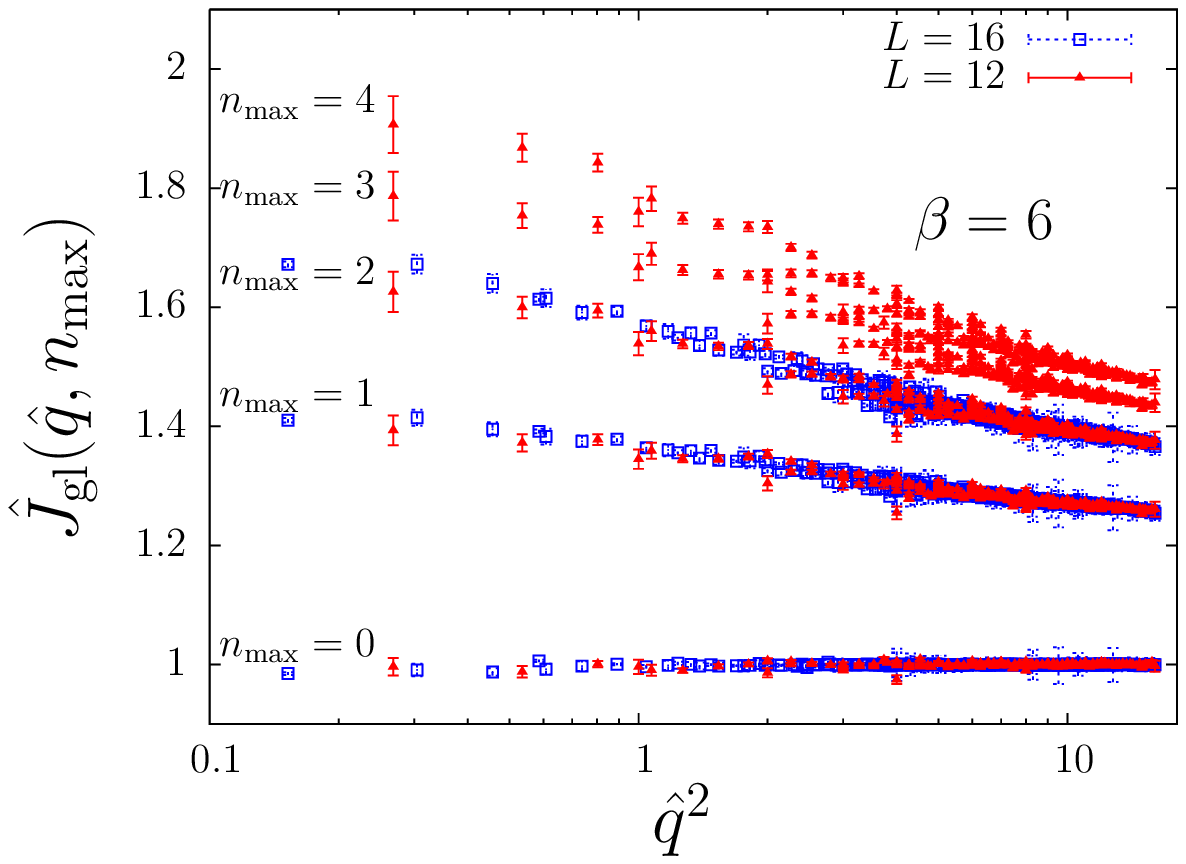}&
    \includegraphics[scale=0.59,clip=true]{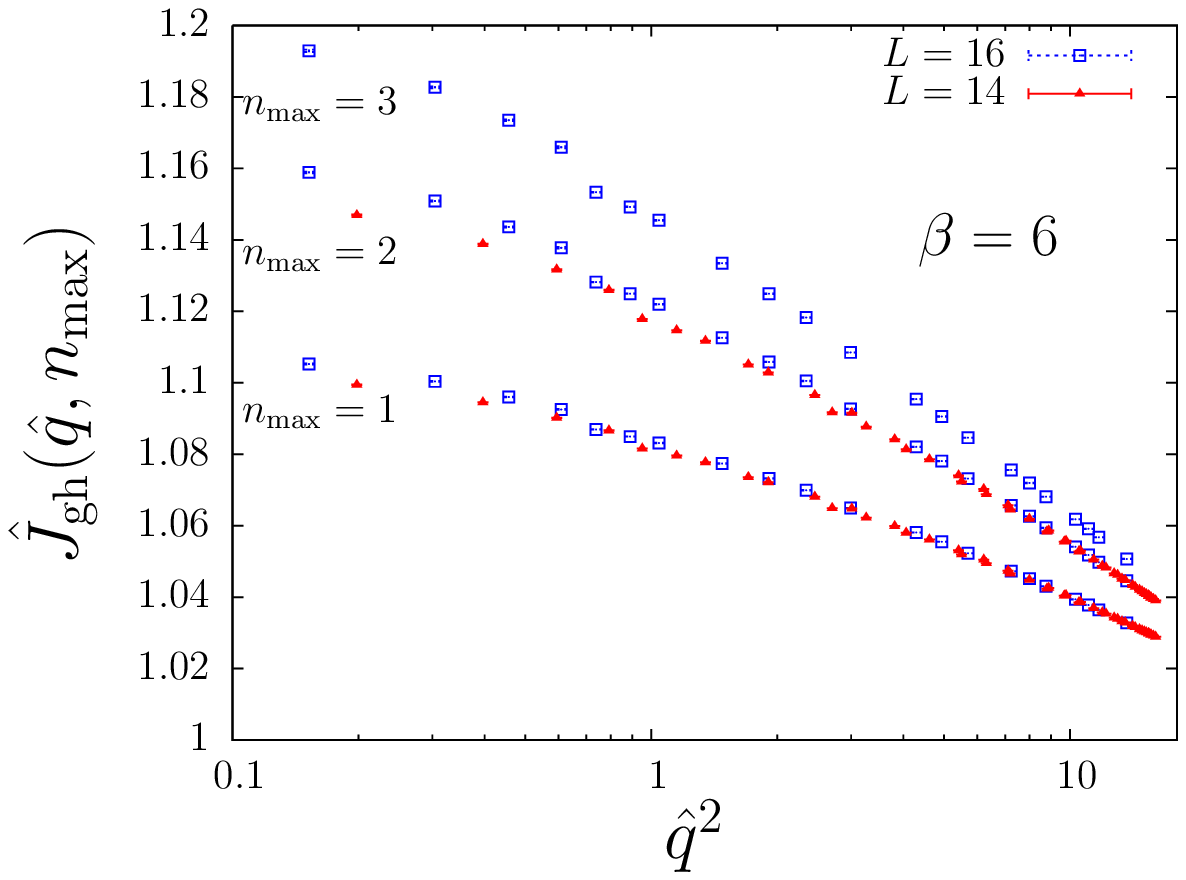}\\
  \end{tabular}
  \vspace{-5mm}
  \caption{Summed gluon (left) and ghost (right) dressing functions for $\beta=6.0$ 
           up to four (left) and three (right) loops for a set of momentum 4-tuples.}
  \label{fig:dress6}
\end{figure}

\begin{figure}[!htb]
  \begin{tabular}{cc}
    \includegraphics[scale=0.59,clip=true]{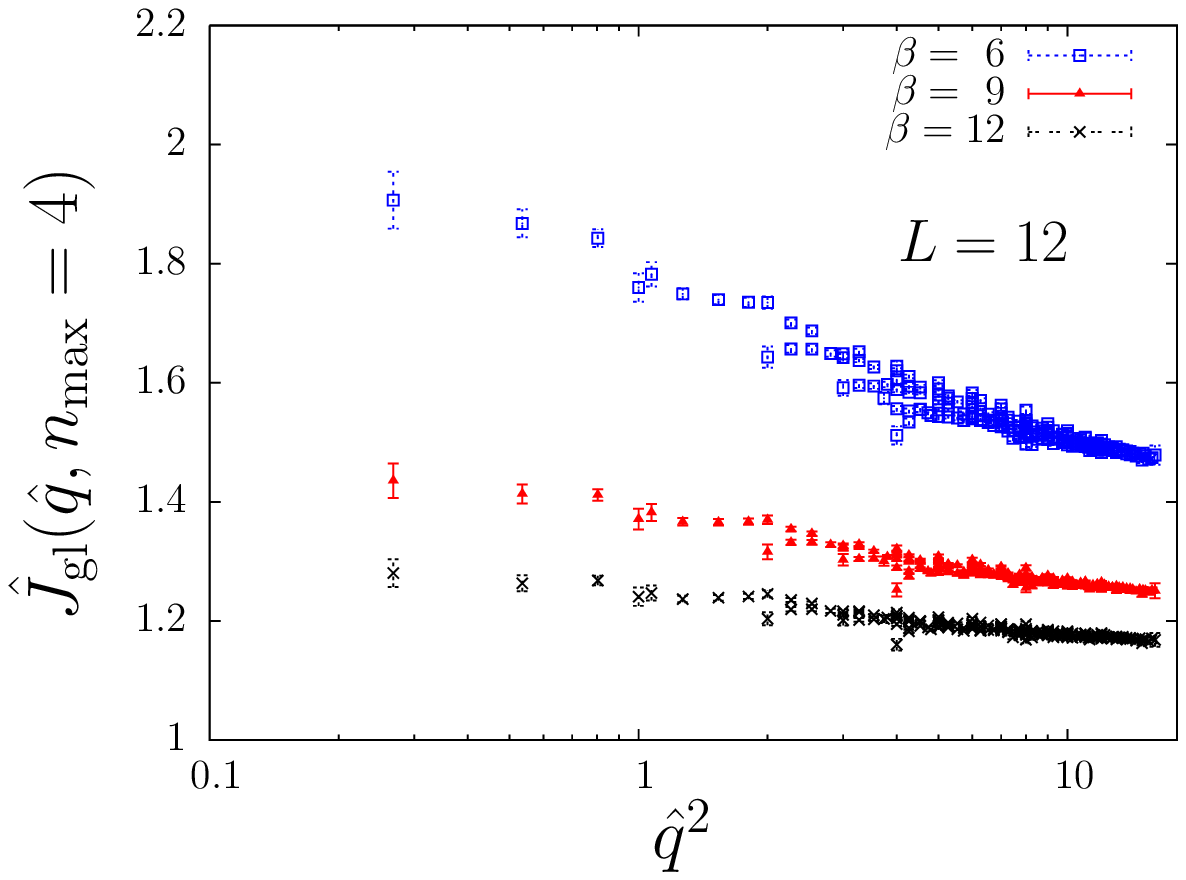}&
    \includegraphics[scale=0.59,clip=true]{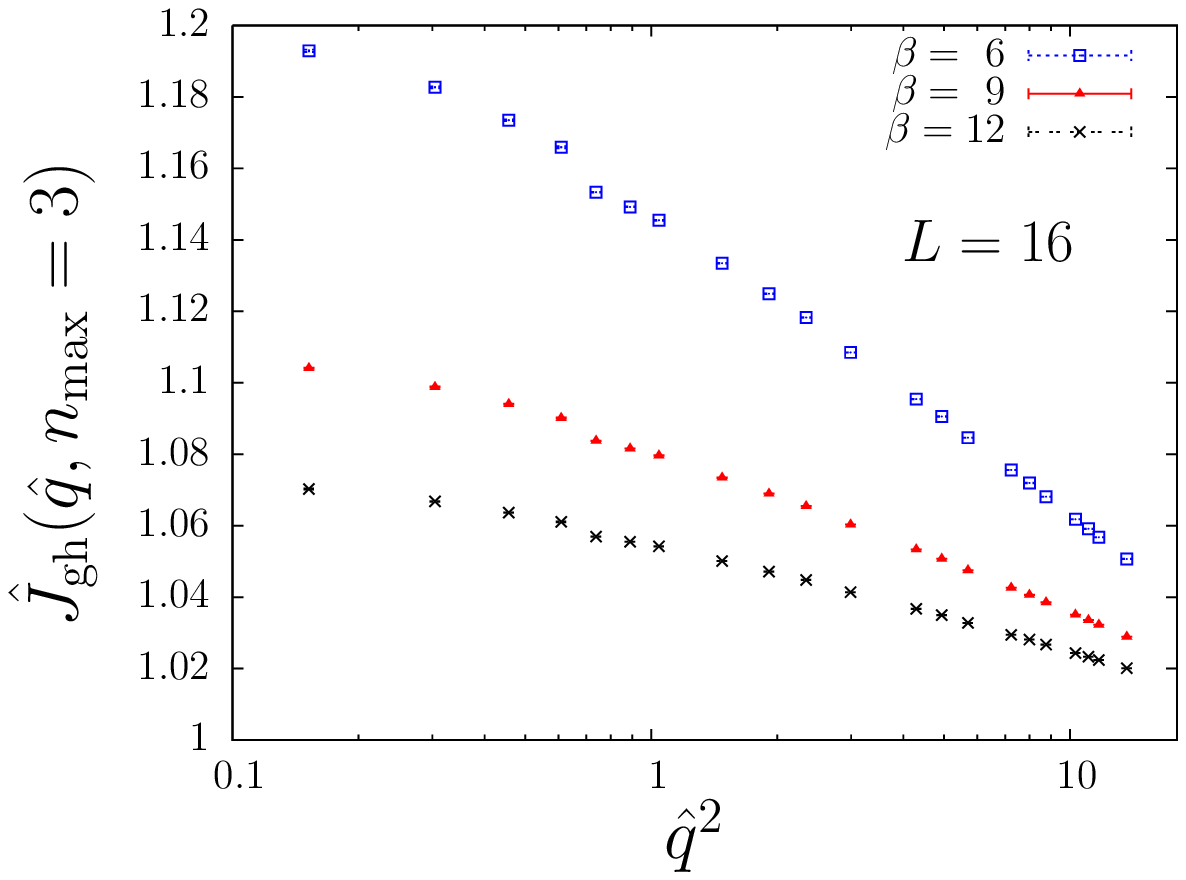}\\
  \end{tabular}
  \vspace{-5mm}
  \caption{Summed perturbative gluon (left) and ghost (right) dressing functions at 
           different $\beta$.}
  \label{fig:dressall}
\end{figure}

The perturbative dressing functions summed to loop order $n_{\rm {max}}$ are calculated 
for a given lattice coupling $\beta$ as follows:
\begin{equation}
  \hat J_{\rm gl/gh}(\hat q,n_{\rm {max}})=\sum_{n=0/1}^{n_{\rm {max}}} \frac{1}{\beta^n}
  \, \hat J_{\rm gl/gh}^{(n)}(\hat q)\,.
\end{equation}
In Figs.~\ref{fig:dress6} we present the summed dressing functions at fixed $\beta$ as function 
of the number $n_{\rm max}$ of loops. Remarkably, all loop contributions are of the same sign, 
such that we expect that these summed-up dressing functions represent a sequence of lower bounds
for the total perturbative function for all momenta $\hat q^2$. 
The variation with the lattice coupling is shown in Figs.~\ref{fig:dressall}
The large loop contributions to the gluon propagator that we find can be interpreted 
as resulting from lattice tadpole contributions.

\section{Summary}
\label{sec:summary}

In NSPT gauge link variables and gauge potentials are naturally related via (\ref{eq:Adef})
different from most nonperturbative numerical implementations.
Using this definition, we have performed higher loop calculations of the lattice gluon and ghost 
propagators in Landau gauge to make predictions for the perturbative content of the propagators 
as function of the lattice momenta taking the hypercubic group into account.

To compare with Monte Carlo data and in this way to find out the nonperturbative contributions 
of the propagators, the logarithmic definition of the gauge fields and the corresponding FP operator 
have to be implemented. Work in that direction is in progress in the Humboldt University group.

A first attempt to estimate the unknown two-loop contribution to the lattice gluon and ghost propagator 
in Landau gauge in the limits of infinite volume and $a q\to 0$ has been presented recently on 
other occasions and can be found in the Proceedings~\cite{Ilgenfritz:2007qj,DiRenzo:2008ir}.

\section*{Acknowledgements}
Part of this work was supported by DFG under contract FOR 465 (Forschergruppe Gitter-Hadronen 
Ph\"anomenologie). E.-M.~I. is grateful to the Karl-Franzens-Universit\"at Graz for the
guest position he presently holds.

\end{document}